# Phases found at grain boundary of $YBa_2Cu_3O_{7-\delta}$ 50 nm films on $SrTiO_3$ by enhanced anomalous scattering at O:K, $Cu:L_{2,3}$ and $Ba:M_{4,5}$ edges

J.V. Acrivos* SJSU, CA 95192-0101

*Dedicated to Nevill Francis Mott, mentor to condensed matter scholars on 100th anniversary of his birth 9/30/1905*

## ABSTRACT

A new phase is detected within 100µm of 24 DEG ab grain boundary (GB) in $YBa_2Cu_3O_{7-\delta}$ 50 nm films on $SrTiO_3$ by enhanced (001) anomalous scattering. Site identification and temperature dependence is interpreted using crystallographic weights to distinguish enhanced scattering from total electron yield and fluorescence spectra. The c-axis, $c_0$ indicates that only ortho-I phase is present far from GB, both ortho-I and II phases are present near GB. The phase $c_0$ is constant versus temperature across the transition to superconductivity.

* jacrivos@athens.sjsu.edu, TEL 408 924 4972, FAX 408 924 4945

## INTRODUCTION

Synchrotron X-ray absorption spectra (XAS) of layered cuprates, $YBCO_{x=6.5\ to\ 6.9}$ where superconducting planes are intercalated between ionic and perhaps magnetic layers are compared at the O:K, $Cu:L_{2,3}$ and the $Ba:M_{4,5}$ edges. The film oxygen composition is obtained from the variation in the c-axis, $c_0$ that determines the (001) enhanced scattered amplitude.

## EXPERIMENTAL

The samples are 50 nm films, grown epitaxially by sputtering in an oxygen atmosphere onto a $SrTiO_3$ crystal with and without a 24 DEG ab grain boundary (GB) at the Complutense University and characterized by synchrotron XRD[1]. Spectra were collected versus photon energy, E at LBNL-ALS 6.3.1 station: by the (001) enhanced scattering ($I_s/I_0$) in the Kortright chamber at different temperatures[2] and distinguished from fluorescence ($F/I_0$) and total electron yield ($TEY/I_0$) in the Nachimuthu chamber where E was calibrated at E(CuO, $Cu:L_3$)=931.2eV[3]. A plane polarized beam (10 by 100 µm wide) of intensity $I_0$, incident on the $1cm^2$ film at position x, at fixed angle θ to the film ab plane (fig. 1) makes an angle 2θ with the detector, as reported in each spectrum (fig. 2-5). The samples are identified by whether the film is deposited on a single or a bi-crystal (SC or BC) qualified by the year fabricated/year measured. The oxygen composition is obtained by the comparison to XRD data[1, 5].

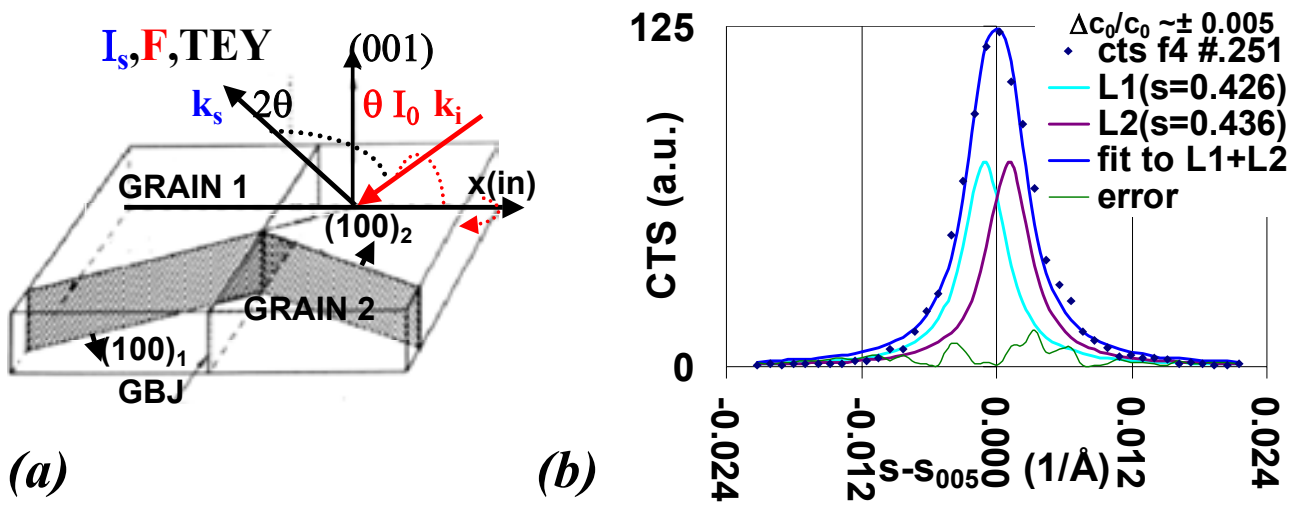

***FIG. 1:*** *Sample: **(a)** Measurement geometry determined by the fixed horizontal incident beam $k_i$, its position and angle θ by the sample displacement and rotation about the* x*-axis, and $k_s$ by the detector angle 2θ to $k_i$. **(b)** BC02/03 (001) XRD versus $s-s_{005} = 2\sin\theta/\lambda - 5/c_0$. 100µm wide beam detected two $c_0$ at GB [1b].*

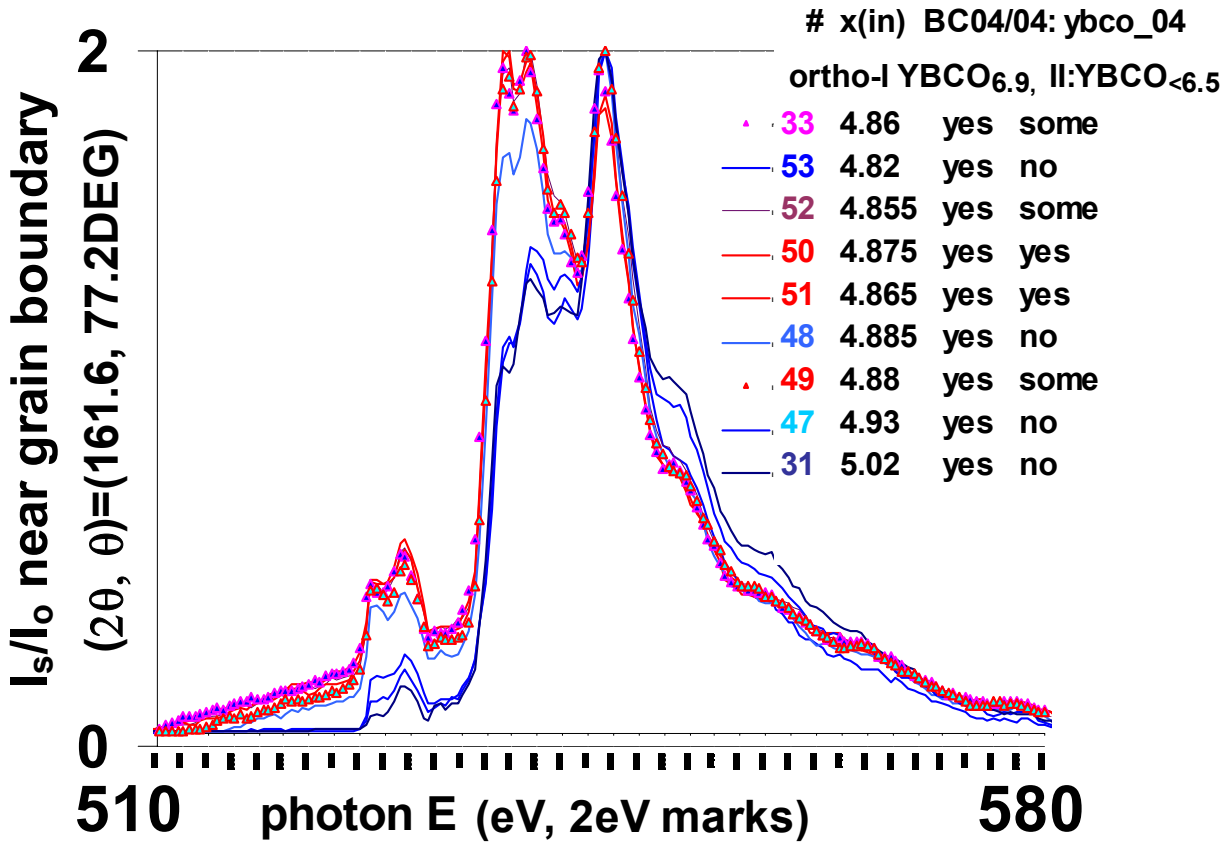

***Figure 2:*** *Phases detected by $I_s/I_0$ as incident beam position x moves across the GB. A new phase is induced within 100µm of GB (Δ, o) detected by enhancement peak at $\Delta E_{Bragg}/\langle E_{Bragg}\rangle \approx -\Delta c_0/c_0$ from the original $c_0$(ortho-I phase)≈ 11.6Å to $c_0$(ortho-II phase)≈11.7Å.*

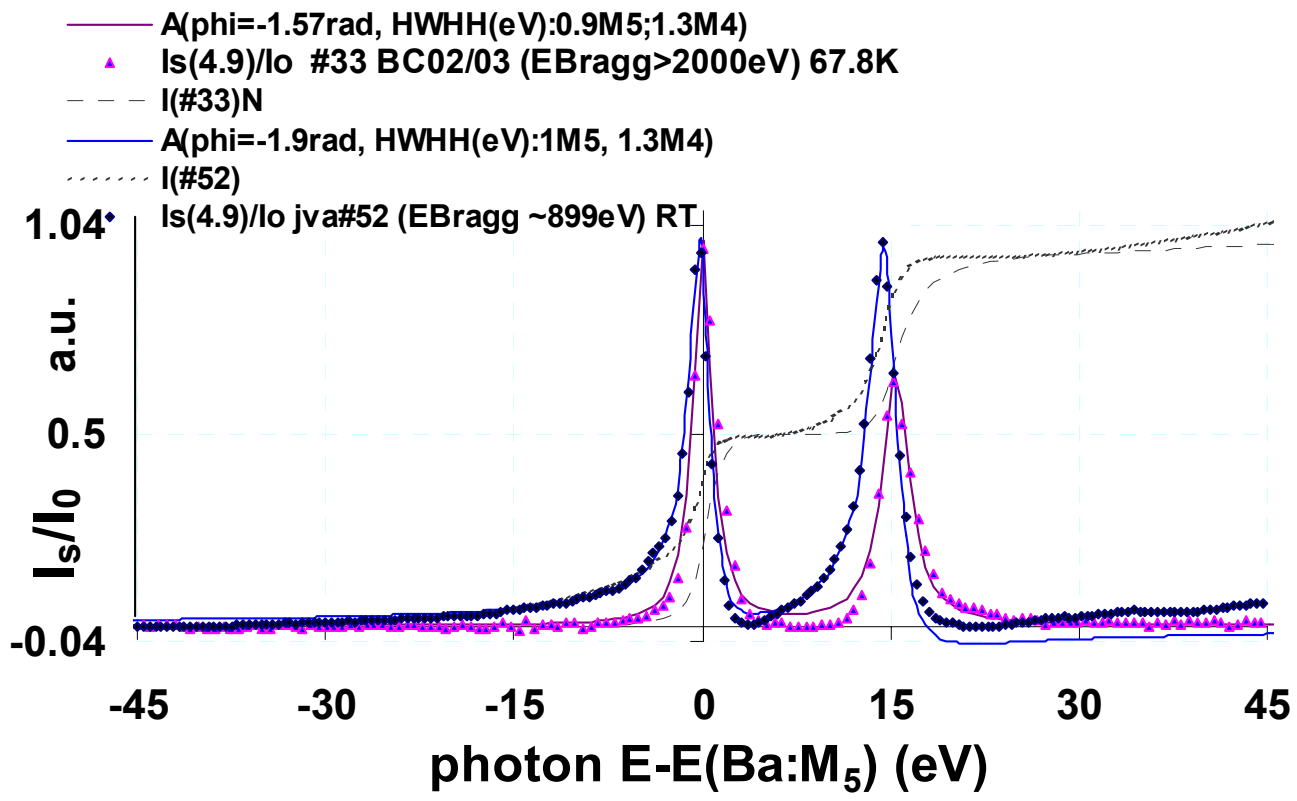

***Figure 3:*** *Effect of $E_{Bragg}$ (2) on $I_s/I_0$ near the $Ba:M_{4,5}$ edges (BC02/03; BC04/04). Lifetime broadening and distortion due to some Ba, commonly occupying Y sites is observed. Broadening is evident in the integrated intensities, I from 730eV and the fit to A (4) with different HWHH at the $M_5$ and $M_4$ WL, but the integrated intensities remain equal even as the lines narrow.*

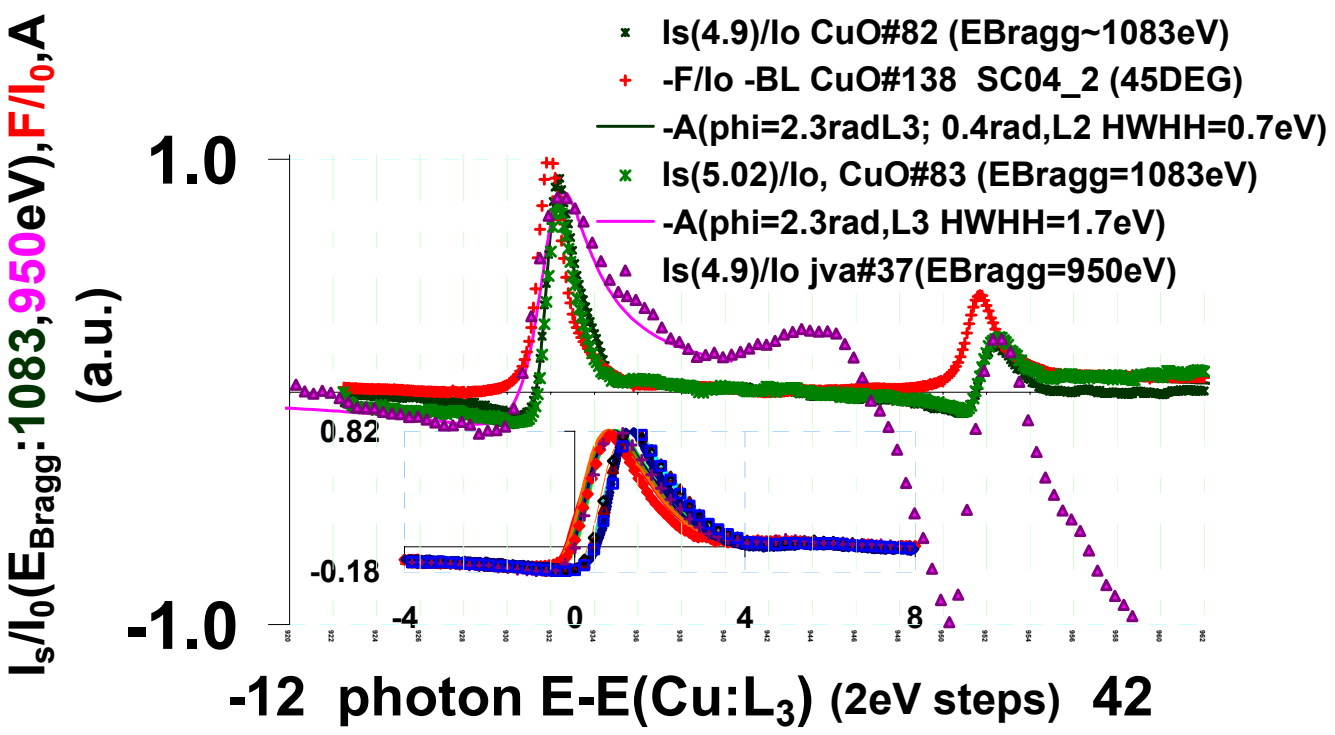

***Figure 4:*** *Effect of $E_{Bragg}$ on $I_s/I_0$ near the Cu:$L_{2,3}$ edges compared to $F/I_0$ and fitted to relation (4) $A(L_3)/A(L_2)=3$, $\alpha_{Cu}$ is constant to 1% in 40eV interval. 100eV broad background at E(Cu:$L_3$) -$E_{Bragg}$ ≈ 18eV disappears at $10^2$eV. Insert shows reversible 24h cycle of $I_s/I_0$ versus T (red $T>T_c$, blue $T<T_c$) with $E_{Bragg}$-E(Cu:L3)>$10^3$eV.*

## DISCUSSION

The YBCO (001) diffraction enhancement is the only one accessible by soft X-rays. The scattering amplitude by atoms j depends on the incident and scattered photon momenta $\mathbf{k_i}, \mathbf{k_s}$ (fig. 1a), polarization $\hat{\mathbf{e}}_i, \hat{\mathbf{e}}_s$ and E [4, 6]:

$$f_j(\mathbf{k_i}, \mathbf{k_s}, E) = f_j^0(\mathbf{k_i}, \mathbf{k_s}) + \Delta f_j(\mathbf{k_i}, \mathbf{k_s}, \hat{\mathbf{e}}_i, \hat{\mathbf{e}}_s, E). \qquad (1)$$

The Thomson amplitude $f_j^0$ is the matrix element of the square of the vector potential acting on the electron number density. The anomalous amplitude:

$$\Delta f_j(\mathbf{k_i}, \mathbf{k_s}, \hat{\mathbf{e}}_i, \hat{\mathbf{e}}_s, E) = f_j' + i f_j'' \approx$$
$$\Sigma_j\Sigma_{nl}[\langle \hat{\mathbf{e}}_s.\mu_{ln}\, e^{-i\mathbf{k_s}.\mathbf{r_j}}\rangle\langle e^{i\mathbf{k_i}.\mathbf{r_j}}\, \mu_{nl}^*.\hat{\mathbf{e}}_i\rangle]/[E_n - E_l + E + (\Delta_n - i\, \mathrm{HWHH})] + \mathrm{HC}$$

involves dipole matrix elements $\mu_{ln}$ between initial and final states (n,l) with energies $E_n$, $E_l$, that depend on orientation in a layer cuprates[7] (incident $\varepsilon_{X\text{-}ray}$ unit vector, $\hat{\mathbf{e}}_i$, is in the film ab plane), state lifetime that determines the half width at half height HWHH, crystallographic site diffraction weights $\alpha_j = \Sigma_j e^{i(\mathbf{k_i}-\mathbf{k_s}).\mathbf{r_j}}$, $\Delta_n$= Lamb shift, $f_j'$= dispersion, $f_j''$= absorption, HC=Hermitean conjugate, and the Bragg relation:

$$E_{Bragg} = hc/2\sin(\theta)/c_0 \;+\; \text{Stenström correction}(\theta) \qquad (2)$$

is determined by the magnitude of the c-axis, $c_0$, h= Planck constant, c= velocity of light. Hanzen[5a] has correlated the oxygen composition of $YBCO_{x=7-\delta}$ with $c_0$ in each phase. Thus at fixed orientation, minute changes:

$$\Delta E_{Bragg}/E_{Bragg} \sim -\Delta c_0/c_0. \qquad (3)$$

can detect the appearance of new phases as the incident beam scans the film surface with fixed $\mathbf{k_i}, \mathbf{k_s}$ (fig. 1a). As E=>$E_{Bragg}$ the edge white line (WL) lifetime broadening increases, due to the detection of enhanced quadrupole transitions and/or enhanced Compton and Rayleigh scattering. As $E_{Bragg}$ -E increases, the tail of $f^0$ becomes a baseline correction (fig. 2-5), but the film may rotate the plane polarized light by an angle $\phi$. Then the signal and its Hilbert-Kramers-Kronig transform:

$$\begin{pmatrix} I_s/I_0\,(E) \\ P\int_{-\infty}^{\infty} \frac{2\, I_s/I_0\,(E')}{\pi\,(E'-E)}\, dE' \end{pmatrix} = \begin{pmatrix} \cos(\phi) & -\sin(\phi) \\ \sin(\phi) & \cos(\phi) \end{pmatrix} \begin{pmatrix} R\Delta f \\ I\Delta f \end{pmatrix}$$

are a mixture of real, R and imaginary, I terms in Δf. The data (fig. 2-5) are analyzed with the purpose to ascertain the properties of films with GB for proper industrial use:

**(i)** A mixture of real and imaginary components is observed in the WL at the Cu:$L_{2,3}$ and Ba:$M_{5,4}$ (fig. 3, 4) for BC and SC films where TEY/$I_0$ and F/$I_0$ show a Lorentzian shaped WL with an edge jump weaker than 1% of WL amplitude[7]. Thus if the $f^0$ tail is linear, the enhanced scattered amplitude minus a base line may be compared to:

$$A_j = I_{sj}/I_0/\alpha_j = [y\cos(\phi) - \sin(\phi)]/[1 + y^2] \qquad (4)$$

where y=(E-$E_0$)/HWHH, $E_0$ is the edge energy and HWHH is the WL half width at half height. The fitted A indicate that the film rotates the plane polarized beam by $\phi$(Cu:$L_3$) ≈ 3π/4±π at E(Cu:$L_3$)-$E_{Bragg}$=18 to $10^2$eV, and f'≈ $I_s/I_0(E_{Bragg})$ (fig. 4) agrees with theory $\phi \approx 0$[4, 6]. The observed lifetime broadening[6b] narrows to HWHH=0.7 from 1.7eV when $E_{Bragg}$-E(Cu:$L_3$)= $10^2$ and 18eV, respectively. WL transitions at $L_{2,3}$ edges (fig. 4) depend on the Cu valence:

$Cu^{1+}(.2p^6..) \Leftrightarrow Cu^{+1}(.2p_{3/2}{}^3..4d_{5/2})$ ; $Cu^{+1}(.2p_{1/2}..4d_{3/2})$,
$Cu^{2+}(.\,2p^6 3d^9) \Leftrightarrow Cu^{+2}(.2p_{3/2}{}^3.3d^{10})$; $Cu^{+2}(.2p_{1/2}..4d_{3/2})$,
$Cu^{+3}(.2p^6 3d^8) \Leftrightarrow Cu^{+3}(.2p_{3/2}{}^3..3d^9)$; $Cu^{+3}(.2p_{1/2}..3d^9)$,

the crystal field splitting (different at the $L_2$ and $L_3$ edges) and orientation[7-9] making it difficult to assign spectral features to the Cu sites in $YBCO_x$. Site identification is made by the variation in $\alpha_j(E) = n_j \cos(2\pi\, z_j\, E/E_{Bragg})$ versus different fixed $E_{Bragg}$, when $n_j$ is the number of equivalent atom j sites with coordinate $z_j$ in the unit cell (Table I). When E(Cu:$L_2$) ≈ $E_{Bragg}$ ≈ 950eV, $\alpha_{Cu:2}/\alpha_{Cu:1}$≈ -1.4 the enhanced shoulders ~ 6eV above the main signal, but of opposite sign amplitude may be due to the Cu:1 site contribution. The exact cancellation expected for $E_{Bragg}$= 1083eV, $\alpha_{Cu:1}/\alpha_{Cu:2}$≈-1, if the second order matrix elements in $\Delta f_{Cu}$ for both sites are of the same order of magnitude, is not observed, indicating that Cu:1 and Cu:2 appear at different E, with a different Cu valence and $\phi$(Cu:2) ≈ 3π/4.

**(ii)** Data at the O:K edge indicate that a displacement of the 100μm wide beam, across the GB detects a new enhancement peak, associated with a higher $c_0$ phase. The relative amplitude (fig. 2, #50, 51) in the XAFS region centered at 538eV ($c_0$≈11.7Å) identifies it with the ortho-II phase ($YBCO_{<6.5}$) relative to that at 546eV ($c_0$≈11.6Å) for the ortho-I phase ($YBCO_{6.9}$) in agreement with the XRD data[1b]. The width of the GB is comparable to the beam width since full enhancement, at $E_{Bragg}$≈ 538eV appears only within x ≈ 4.87±0.005in, while that at 546eV decreases very little across the GB. Comparison to $c_0$ data versus O composition[5] indicates that near the GB a ~5% discontinuous O decrease induces the ortho-II phase which releases the film strain, by creating the $k_x$= -$k_y$ periodic lattice distortions (PLD) observed in XRD for the film[1b,c, 8c].

The information obtained on $YBCO_x$ transitions:

$O{:}1s^2. \Leftrightarrow O{:}1s.np_{x,y}$, n>2 and $O{:}1s^2. \Leftrightarrow O{:}1s.np_z$, n>2

when $\hat{\mathbf{e}}_i$ is in the ab plane is according to relation (1). Comparison of enhancement at $E_{Bragg}$= 545 and 2*$10^3$eV (Table I, fig. 5 #37, 28) identifies the site contributions. $I_s/I_0$

near 528, 538eV is identified with O:2 by the doubling of α(O:2) and orientation independent amplitude maximum in $F/I_0$, expected in a nearly local octahedral field. $I_s/I_0$ near 530eV is identified with O:1 by the relative amplitudes α(O:2)/α(O:1)= 0.8 and 1.9, and $I_s/I_0$ near 531eV to 536eV is identified with $O{:}3_{A,B}$ by $\alpha(O{:}3_{A,B})$ sign changes:

$$I_s/I_0(\alpha(O{:}3_{A,B})\approx -2.7) \leq 0 \text{ while } I_s/I_0(\alpha(O{:}3_{A,B})\approx 3.5) > 0.$$

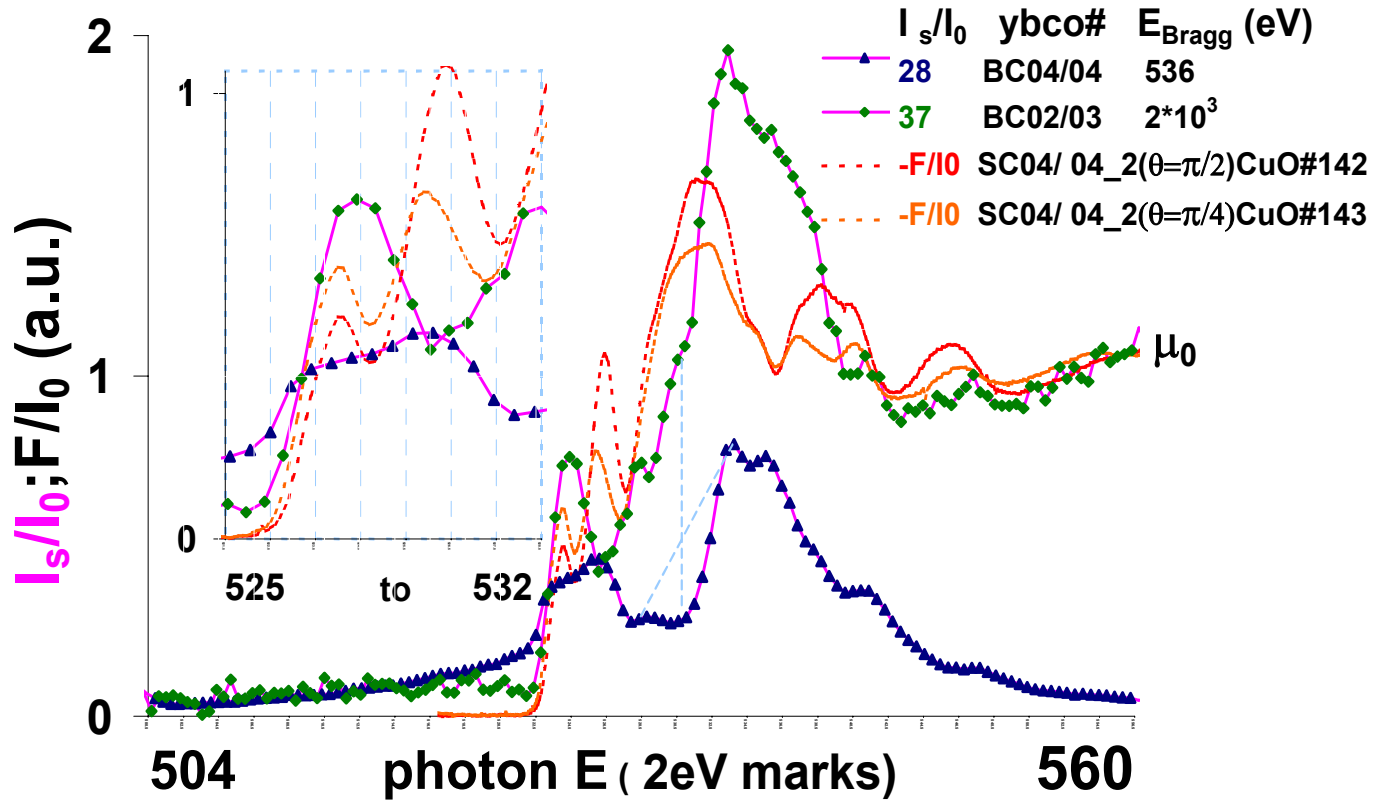

***Figure 5:*** *O:K edge comparison of $I_s/I_0(E_{Bragg}\approx$ 536 and $2*10^3$eV) with $F/I_0$, normalized to XAFS: $\mu_0 =1$, and $I_s/I_0$ to constant α(O:1) at~530eV identify O:2 and $O{:}3_{AB}$ site contributions by the variation in α versus $E_{Bragg}$ (Table I).*

Assignments made by a single diffraction with soft X-rays using the variation of α versus $E_{Bragg}$ are similar to those made by different measurements[8] for de-twinned single crystals $YBCO_x$, and may be correlated to the chemical valence: The most negative ionic valence is associated with the lowest energy for site O:2 in the BaO layer, the next higher energy with site O:1 in the CuO chains and the highest valence is assigned to sites $O{:}3_{A,B}$ where molecular orbital calculations show that the $CuO_2$ layer in $YBCO_x$ nano-particles is covalent with a Mulliken atomic charge at the $O{:}3_{A,B}$ sites of –1.3 and 0.8 at the Cu:2 sites[1c, 9].

**Table I:** Assignment of the $YBCO_x$ unit cell sites, Hanzen notation[5a] by correlation of $I_s/I_0$ to crystallographic diffraction weights α when $E_= = E_{Bragg}$ and $E_{\neq} \neq E_{Bragg}$ (fig. 2- 5).

| $YBCO_x$ | $D^{17}_{2h}$ | Site: | O:1 | O:2 | O:3A,B | Cu:1 | Cu:2 | Ba | Y |
|---|---|---|---|---|---|---|---|---|---|
| (001) enhancement | | z: | 0 | ± .18 | ± .378 | 0 | ± .358 | ± .18 | ± .5 |
| | α(E=): | | 1 | 0.8 | -2.9 | 1 | -1.4 | 0.8 | -1 |
| $I_s/I_0$ Peak | | α(E≠): | 1 | 1.9 | 3.4 | 1 | –1 | 1.9 | 0.6 |
| E (eV) | Sign $I_s/I_0$ | | Site Contribution to Signal | | | | | | |
| 528 | + | + | | yes | | | | | |
| 530 | + | | yes | | | | | | |
| 531-535 | - | | | | yes | | | | |
| 531-535 | | + | | | yes | | | | |
| 538 | + | | maybe | yes | | | | | |
| 545 | + | | | | yes | | | | |
| 779 | + | + | | | | | | yes | some Ba |
| 793 | + | + | | | | | | yes | some Ba |
| 932 | ± | ± | | | | | yes | | |
| 950 | -± | ± | | | | | yes | | |
| 936 | -± | | | | | yes | | | |
| 954 | -± | | | | | yes | | | |

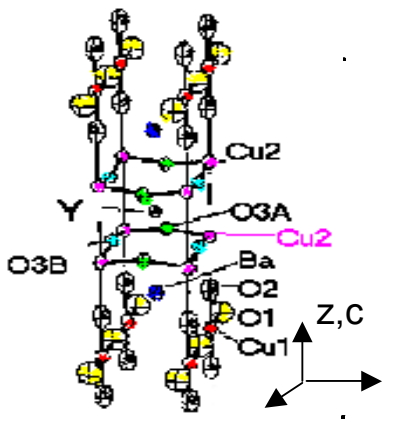

Temperature dependence measurements indicate that $c_0$ is unchanged across $T_c$ by the constant enhancement peak observed at $E_{Bragg}\approx$ 546eV in the ortho-I phase. $\phi(Cu{:}2{:}L_{3,2}) \approx 3\pi/4$ in (4) is constant across $T_c$, but a reversible 0.5eV edge shift below $T_c$, observed at the $Cu{:}2{:}L_{3,2}$ edges in a 24 h, T cycle (fig. 4 insert) is assigned to an increased Cu:2 site valence below $T_c$.

**(iii)** In the ideal $YBCO_x$, Ba occupies a unique site but in real crystals it also occupies Y sites. As $E_{Bragg}$ - $E(Ba{:}M_5)$= -38 to $2*10^3$eV lifetime broadening is observed when $\alpha_{Ba}/\alpha_Y\approx$ -0.8 but the WL narrow for $\alpha_{Ba}/\alpha_Y \approx 3$ (Table I, fig. 3 #52 and #33). Amplitudes for WL transitions:

$$Ba^{2+}(.3d^{10}.) \Leftrightarrow Ba^{+2}(.3d_{5/2}{}^{5}.4f_{7/2});\ Ba^{+2}(.3d_{3/2}{}^{3}.4f_{5/2}),$$

are proportional to the initial state multiplicities, $A_{M5}/A_{M4}$ = 1.5 only for $E_{Bragg}$-E> $2*10^3$eV. The $TEY/I_0$ for BC04/04 are orientation dependent[7c, 9].

## CONCLUSION

The fabricated nano-film $YBCO_x$ (001) anomalous enhanced scattering analysis, sensitive to phase and O composition characterizes the GB for device applications and theoretical interpretation of transport data.

## ACKNOWLEDGEMEMNTS

Work was supported by the NSF and Dreyfus Foundations at SJSU, DOE at LBNL-ALS and scholars[1-9].

## REFERENCES

[1](a) M. A. Navacerrada, M. L. Lucía and F. Sánchez-Quesada, *Europhys. Lett* **54**, 387 (2001); (b) M.A. Navacerrada and J.V. Acrivos, *NanoTech2003*, **1**, 751 (2003); (c) H.S. Sahibudeen, J.V. Acrivos and M.A. Navacerrada, *Nanotech 2005*, **2**, 573 (2005)

[2] J.B. Kortright et al., *J. Magn. Magn. Materials*, **207**, 7 (1999)

[3]P.Nachimuthu et al., *Chem. Mater.* **15**, 3939 (2003)

[4] (a) R.W. James, "*The optical principles of X-rays*", Ox Bow Press, Woodbridge, Conn. 1962; (b) L.K. Templeton and D.H. Templeton, *Acta Cryst.* **A47**, 414 (1991)

[5](a) R.M. Hanzen et al., "*Physical Properties of High Temperature Superconductors, II*" p. 121, D.M. Ginsberg, ed, World Scientific, Singapore (1990); (b) N.H. Andersen, *Physica C,* **317-318**, 259 (1999)

[6] (a) L.B. Sorensen et al., *"Diffraction anomalous fine structure"* North Holland, G. Materlik et al., ed (1994) p.389; (b) Hämäläinen, et al., *ibid* p. 48; (c) J.C. Woicik et al., *Phys. Rev.B* **58,** R4215 (1998)

[7](a) A. Bianconi et al, *Phys. Rev.B* **38,** 7196 (1988); (b) F.M.F. de Groot et al., *ibid. B* **42,** 5459 (1990); (c) J.V. Acrivos et al., arKiv, cond-mat/0504369

[8](a) M. Merz, et al., *Phys. Rev. Lett.* **80**, 5192 (1998); (b) J.H. Guo et al., *Phys. Rev. B* **61**, 9140 (2000); (c) E. Bascones et al., *ibid*, **71,** 012505 (2005)

[9](a) J.V. Acrivos, *Solid State Sciences*, **2**, 807 (2000); (b) J.V. Acrivos et al., *Microchemical Journal,* **71**, 117 (2002); *ibid,* **81**, 98 (2005)